# AWESOME: A General Multiagent Learning Algorithm that Converges in Self-Play and Learns a Best Response Against Stationary Opponents

**Vincent Conitzer**　　　　　　　　　　　　　　　　　　　　　　　　　　　　CONITZER@CS.CMU.EDU
Computer Science Department, Carnegie Mellon University, 5000 Forbes Avenue, Pittsburgh, PA 15213

**Tuomas Sandholm**　　　　　　　　　　　　　　　　　　　　　　　　　　　　SANDHOLM@CS.CMU.EDU
Computer Science Department, Carnegie Mellon University, 5000 Forbes Avenue, Pittsburgh, PA 15213

## Abstract

A satisfactory multiagent learning algorithm should, *at a minimum*, learn to play optimally against stationary opponents and converge to a Nash equilibrium in self-play. The algorithm that has come closest, WoLF-IGA, has been proven to have these two properties in 2-player 2-action repeated games—assuming that the opponent's (mixed) strategy is observable. In this paper we present AWESOME, the first algorithm that is guaranteed to have these two properties in *all* repeated (finite) games. It requires only that the other players' actual actions (not their strategies) can be observed at each step. It also learns to play optimally against opponents that *eventually become* stationary. The basic idea behind AWESOME (*Adapt When Everybody is Stationary, Otherwise Move to Equilibrium*) is to try to adapt to the others' strategies when they appear stationary, but otherwise to retreat to a precomputed equilibrium strategy. The techniques used to prove the properties of AWESOME are fundamentally different from those used for previous algorithms, and may help in analyzing other multiagent learning algorithms also.[1]

## 1. Introduction

Learning from experience is a key capability in AI, because it can be difficult to program a system in advance to act appropriately. Learning is especially important in multiagent settings where the other agents' behavior is not known in advance. Multiagent learning is complicated by the fact that the other agents may be learning as well, thus making the environment nonstationary for a learner.

Multiagent learning has been studied with different objectives and different restrictions on the game and on what the learner can observe (e.g., (Tan, 1993; Sen & Weiss, 1998)). Two *minimal* desirable properties of a good multiagent learning algorithm are

• Learning to play optimally against stationary opponents (or even opponents that eventually become stationary).[2]

• Convergence to a Nash equilibrium in self-play (that is, when all the agents use the same learning algorithm).

These desiderata are *minimal* in the sense that any multiagent learning algorithm that fails at least one of these properties is, in a sense, unsatisfactory. Of course, one might also want the algorithm to have additional properties.[3]

---

[1] We thank the anonymous reviewers, as well as Michael Bowling and Manuela Veloso for helpful discussions. This material is based upon work supported by the National Science Foundation under CAREER Award IRI-9703122, Grant IIS-9800994, ITR IIS-0081246, and ITR IIS-0121678.

[2] This property has sometimes been called *rationality* (Bowling & Veloso, 2002), but we avoid that term because it has an established, different meaning in economics.

[3] It can be argued that the two properties are not even strong enough to constitute a "minimal" set of requirements, in the sense that we would still not necessarily be satisfied with an algorithm if it has these properties. However, we would likely not be satisfied with any algorithm that did *not* meet these two requirements, even if it had

However, to date there has been no algorithm that achieves both of these minimal properties in general repeated games. Many of the proposed algorithms satisfy the first property (e.g. (Vrieze, 1987; Claus & Boutilier, 1998; Singh et al., 2000; Bowling & Veloso, 2002; Wang & Sandholm, 2002)). Some of the algorithms satisfy the second property in restricted games (e.g. (Vrieze, 1987; Littman, 1994; Hu & Wellman, 1998; Singh et al., 2000; Bowling & Veloso, 2002; Wang & Sandholm, 2002)).

The algorithm that has come closest to satisfying both of the properties in general repeated games is WoLF-IGA (Bowling & Veloso, 2002). (This algorithm set out to do exactly this and was an improvement over an earlier algorithm (Singh et al., 2000).) It is guaranteed to have both of the properties in general games under the following assumptions: (a) there are at most 2 players, (b) each player has at most 2 actions to choose from, (c) the opponent's strategy (distribution over actions) is observable, and (d) gradient ascent of infinitesimally small step sizes can be used.[4]

Another learning algorithm that succeeds in achieving similar goals is "regret matching", with which the learner's regrets converge to zero and, if all players use the learning algorithm, the empirical distributions of play converge to a *correlated* equilibrium (Hart & Mas-Colell, 2000). (The set of correlated equilibria is a strict superset of the set of Nash equilibria, where players are allowed to condition their action on a commonly observed signal. Thus, convergence to a Nash equilibrium is a strictly stronger property than convergence to a correlated equilibrium.) Convergence to correlated equilibria is achieved by a number of other learning procedures (Cahn, 2000; Foster & Vohra, 1997; Fudenberg & Levine, 1999).

In this paper we present AWESOME, the first algorithm that has both of the desirable properties in general repeated games.[5] It removes all of the assumptions (a)–(d). It has the two desirable properties with any finite number of agents and any finite number of actions; it only requires being able to observe other players' actions (rather than the distribution that the actions are drawn from); and it does not rely on infinitesimal updates.

AWESOME still makes some of the same assumptions that were made in the prior theoretical work attempting to attain both of the desirable properties (Singh et al., 2000; Bowling & Veloso, 2002). First, (for now) it only deals with repeated games—that is, stochastic games with a single state. Second, it assumes that the structure of the game is known (has already been learned). Third, as in (Bowling & Veloso, 2002), we also assume that the agents can compute a Nash equilibrium.[6] (It is still unknown whether a Nash equilibrium can be found in worst-case polynomial time (Papadimitriou, 2001), but it is known that certain related questions are hard in the worst case (Conitzer & Sandholm, 2003).)[7]

The basic idea behind AWESOME (*Adapt When Everybody is Stationary, Otherwise Move to Equilibrium*) is to try to adapt to the other agents' strategies when they appear stationary, but otherwise to retreat to a precomputed equilibrium strategy. At any point in time, AWESOME maintains either of two null hypotheses: that the others are playing the precomputed equilibrium, or that the others are stationary. Whenever both of these hypotheses are rejected, AWESOME restarts completely. AWESOME may reject either of these hypotheses based on actions played in an epoch. Over time, the epoch length is carefully increased and the criterion for hypothesis rejection tightened to obtain the convergence guarantee. The AWESOME algorithm is also self-aware: when it detects that its own actions signal nonstationarity to the others, it restarts itself for synchronization purposes.

The techniques used in proving the properties of AWESOME are fundamentally different from those used for previous algorithms, because the requirement that the opponents' whole strategies can be observed is dropped. These techniques may also be valuable in the analysis of other learning algorithms in games.

It is important to emphasize that, when attempting to converge to an equilibrium, as is common in the literature, our goal is to eventually learn the equilibrium of the *one-shot* game, which, when played repeatedly, will also constitute an equilibrium of the repeated

---

other properties. This is the sense in which we use the word "minimal".

[4]Bowling and Veloso also defined a more generally applicable algorithm based on the same idea, but only gave experimental justification for it.

[5]As in WoLF-IGA, our notion of convergence is that the *stage-game* strategy converges to the desired strategy (not just the long-term empirical distribution).

[6]We assume that when there are multiple AWESOME players, they compute the same Nash equilibrium. This is natural since they share the same algorithm.

[7]Much of the literature on learning in games is concerned with learning (the relevant aspects of) the game itself, or, even when the game is already known, with reaching the equilibrium through some simple dynamics (not using a separate algorithm to compute it). These are certainly worthwhile objectives in our opinion. However, in this paper the uncertainty stems from the opponent, and the goal is to play appropriately with respect to the opponent's algorithm.

game. The advantage of such equilibria is that they are natural and simple, always exist, and are robust to changes in the discounting/averaging schemes. Nevertheless, in repeated games it is possible to also have equilibria that are fundamentally different from repetitions of the one-shot equilibrium; such equilibria rely on a player conditioning its future behavior on the opponents' current behavior. Interestingly, a recent paper shows that when players are interested in their average payoffs, such equilibria can be constructed in worst-case polynomial time (Littman & Stone, 2003).

The rest of the paper is organized as follows. In Section 2, we define the setting. In Section 3, we motivate and define the AWESOME algorithm and show how to set its parameters soundly. In Section 4, we show that AWESOME converges to a best response against opponents that (eventually) play stationary strategies. In Section 5, we show that AWESOME converges to a Nash equilibrium in self-play. In Sections 6 and 7, we present conclusions and directions for future research.

## 2. Model and definitions

We study multiagent learning in a setting where a fixed finite number of agents play the same finite stage game repeatedly. We first define the stage game and then the repeated game.

### 2.1. The stage game

**Definition 1 (Stage game)** *A* stage game *is defined by a finite set of agents* $\{1, 2, \ldots, n\}$, *and for each agent $i$, a finite action set $A_i$, and a utility function $u_i : A_1 \times A_2 \times \ldots \times A_n \to \mathbb{R}$. The agents choose their actions independently and concurrently.*

We now define strategies for a stage game.

**Definition 2 (Strategy)** *A* strategy *for agent $i$ (in the stage game) is a probability distribution $\pi_i$ over its action set $A_i$, indicating what the probability is that the agent will play each action. In a* pure strategy, *all the probability mass is on one action. Strategies that are not pure are called* mixed strategies.

The agents' strategies are said to be in equilibrium if no agent is motivated to unilaterally change its strategy given the others' strategies:

**Definition 3 (Nash equilibrium (NE))** *A strategy profile $(\pi_1^*, \pi_2^*, \ldots, \pi_n^*)$ is a* Nash equilibrium (NE) *(of the stage game) if, for every agent $i$ and for any strategy $\pi_i$,*

$$E_{(\pi_1^*, \ldots, \pi_{i-1}^*, \pi_i^*, \pi_{i+1}^*, \pi_2^*, \ldots, \pi_n^*)} u_i(a_1, a_2, \ldots, a_n) \geq$$
$$E_{(\pi_1^*, \ldots, \pi_{i-1}^*, \pi_i, \pi_{i+1}^*, \pi_2^*, \ldots, \pi_n^*)} u_i(a_1, a_2, \ldots, a_n)$$

*We call a NE a* pure-strategy NE *if all the individuals' strategies in it are pure. Otherwise, we call it a* mixed-strategy NE.

As in most of the game theory literature on learning (for a review, see (Fudenberg & Levine, 1998)) and in both of the theoretical results on multiagent learning in computer science that we are trying to improve upon (Bowling & Veloso, 2002; Singh et al., 2000), we assume that the agents know the game. So, they do not need to learn what the game is, but rather they just need to learn how to play.[8]

### 2.2. The repeated game

The agents play the stage game repeatedly (forever). As usual, we assume that the agents observe each others' actions. An agent may learn from previous rounds, so its strategy in a stage game may depend on how the earlier stage games have been played.

In the next section we present our learning algorithm for this setting, which has the desirable properties that it learns a best response strategy againts opponents that eventually are stationary, and it converges to a Nash equilibrium in self-play.

## 3. The AWESOME algorithm

In this section we present the AWESOME algorithm. We first give the high-level idea, and discuss some additional specifications and their motivation. We then give the actual algorithm and the space of valid parameter vectors for it.

### 3.1. The high-level idea

Roughly, the idea of the algorithm is the following. When the others appear to be playing stationary strategies, AWESOME adapts to play the best response to those apparent strategies. When the others appear to be adapting their strategies, AWESOME retreats to an equilibrium strategy. (Hence, AWESOME stands for *Adapt When Everybody is Stationary, Otherwise Move to Equilibrium.*)

---

[8]If the game were not known and the agents are using the same learning algorithm, they could explore the game and learn the game structure, and then learn how to play. This is also an active area of research in multiagent systems in computer science (e.g., (Littman, 1994; Hu & Wellman, 1998; Claus & Boutilier, 1998; Wang & Sandholm, 2002)).

## 3.2. Additional specifications

While the basic idea is simple, we need a few more technical specifications to enable us to prove the desired properties.

- To make the algorithm well-specified, we need to specify which equilibrium strategy AWESOME retreats to. We let AWESOME compute an equilibrium in the beginning, and it will retreat to its strategy in that equilibrium every time it retreats. To obtain our guarantee of convergence in self-play, we also specify that each AWESOME agent computes the same equilibrium (this is reasonable since they share the same algorithm). We observe that *any* equilibrium will work here (e.g., a social welfare maximizing one), but AWESOME might not converge to *that* equilibrium in self-play.

- We specify that when retreating to the equilibrium strategy, AWESOME forgets everything it has learned. So, retreating to an equilibrium is a complete restart. (This may be wasteful in practice, but makes the analysis easier.)

- To avoid nonconvergence in self-play situations where best-responding to strategies that are not quite the precomputed equilibrium strategies would lead to rapid divergence from the equilibrium, AWESOME at various stages has a null hypothesis that the others are playing the precomputed equilibrium. AWESOME will not reject this hypothesis unless presented with significant evidence to the contrary.

- AWESOME rejects the equilibrium hypothesis also when its own actions, chosen according to its mixed equilibrium strategy, happen to appear to indicate a nonequilibrium strategy (even though the underlying mixed strategy is actually the equilibrium strategy). This will help in proving convergence in self-play by making the learning process synchronized across all AWESOME players. (Since the other AWESOME players will restart when they detect such nonstationarity, this agent restarts itself to stay synchronized with the others.)

- After AWESOME rejects the equilibrium hypothesis, it randomly picks an action and changes its strategy to always playing this action. At the end of an epoch, if another action would perform *significantly* better than this action against the strategies the others appeared to play in the last epoch, it switches to this action. (The significant difference is necessary to prevent the AWESOME player from switching back and forth between multiple best responses to the actual strategies played.)

- Because the others' strategies are unobservable (only their actions are observable), we need to specify how an AWESOME agent can reject, based on others' actions, the hypothesis that the others are playing the precomputed equilibrium strategies. Furthermore, we need to specify how an AWESOME agent can reject, based on others' actions, the hypothesis that the others are drawing their actions according to stationary (mixed) strategies. We present these specifications in the next subsection.

## 3.3. Verifying whether others are playing the precomputed equilibrium and detecting nonstationarity

Let us now discuss the problem of how to reject, based on observing the others' actions, the hypothesis that the others are playing according to the precomputed equilibrium strategies. AWESOME proceeds in epochs: at the end of each epoch, for each agent $i$ in turn (including itself), it compares the actual distribution, $h_i$, of the actions that $i$ played in the epoch (i.e. what percentage of the time each action was played) against the (mixed) strategy $\pi_i^*$ from the precomputed equilibrium. AWESOME concludes that the actions are drawn from the equilibrium strategy if and only if the distance between the two distributions is small: $\max_{a_i \in A_i} |p_{h_i}^{a_i} - p_{\pi_i^*}^{a_i}| < \epsilon_e$, where $p_\phi^a$ is the percentage of time that action $a$ is played in $\phi$.

When detecting whether or not an agent is playing a stationary (potentially mixed) strategy, AWESOME uses the same idea, except that in the closeness measure, in place of $\pi_i^*$ it uses the actual distribution, $h_i^{prev}$, of actions played in the epoch just preceding the epoch that just ended. Also, a different threshold may be used: $\epsilon_s$ in place of $\epsilon_e$. So, AWESOME maintains the stationarity hypothesis if and only $\max_{a_i \in A_i} |p_{h_i}^{a_i} - p_{h_i^{prev}}^{a_i}| < \epsilon_s$.

The naive implementation of this keeps the number of iterations $N$ in each epoch constant, as well as $\epsilon_e$ and $\epsilon_s$. Two problems are associated with this naive approach. First, even if the actions are actually drawn from the equilibrium distribution (or a stationary distribution when we are trying to ascertain stationarity), there is a fixed nonzero probability that the actions taken in any given epoch, by chance, do not appear to be drawn from the equilibrium distribution (or, when ascertaining stationarity, that the actual distributions of actions played in consecutive epochs do not look alike).[9] Thus, with probability 1, AWESOME would eventually restart. So, AWESOME could never converge (because it will play a random action between each pair of restarts). Second, AWESOME would not be able to distinguish a strategy from the precomputed equilibrium strategy if those strategies are within $\epsilon_e$ of each other. (Similarly, AWESOME would not be able to detect nonstationarity if the distributions of actions played in consecutive epochs are within $\epsilon_s$.)

---
[9]This holds for all distributions except those that correspond to pure strategies.

We can fix both these problems by letting the distance $\epsilon_e$ and $\epsilon_s$ decrease each epoch, while simultaneously increasing the epoch length $N$. If we increase $N$ sufficiently fast, the probability that the equilibrium distribution would by chance produce a sequence of actions that does not appear to be drawn from it will decrease each epoch in spite of the decrease in $\epsilon_e$. (Similarly, the probability that a stationary distribution will, in consecutive epochs, produce action distributions that are further than $\epsilon_s$ apart will decrease in spite of the decrease in $\epsilon_s$.) In fact, these probabilities can be decreased so fast that there is nonzero probability that the equilibrium hypothesis (resp. stationarity hypothesis) will *never* be rejected over an infinite number of epochs. Chebyshev's inequality, which states that $P(|X - E(X)| \geq t) \leq \frac{Var(X)}{t^2}$, will be a crucial tool in demonstrating this.

### 3.4. The algorithm skeleton

We now present the backbone of the algorithm for repeated games.

First we describe the variables used in the algorithm. $Me$ refers to the AWESOME player. $\pi_p^*$ is player $p$'s equilibrium strategy. $\phi$ is the AWESOME player's current strategy. $h_p^{prev}$ and $h_p^{curr}$ are the histories of actions played by player $p$ in the previous epoch and the epoch just played, respectively. ($h_{-Me}^{curr}$ is the vector of all $h_p^{curr}$ besides the AWESOME player's.) $t$ is the current epoch (reset to 0 every restart). $APPE$ (all players playing equilibrium) is true if the equilibrium hypothesis has not been rejected. $APS$ (all players stationary) is true if the stationarity hypothesis has not been rejected. $\delta$ is true if the equilibrium hypothesis was just rejected (and gives one round to adapt before the stationarity hypothesis can be rejected). $\epsilon_e^t, \epsilon_s^t, N^t$ are the values of those variables for epoch $t$. $n$ is the number of players, $|A|$ the maximum number of actions for a single player, $\mu$ (also a constant) the utility difference between the AWESOME player's best and worst outcomes in the game.

Now the functions. ComputeEquilibriumStrategy computes the equilibrium strategy for a player. Play takes a strategy as input, and plays an action drawn from that distribution. Distance computes the distance (as defined above) between strategies (or histories). V computes the expected utility of playing a given strategy or action against a given strategy profile for the others.

We are now ready to present the algorithm.

AWESOME()
1. **for each** $p$
2. $\pi_p^* := \text{ComputeEquilibriumStrategy}(p)$
3. **repeat** {// beginning of each restart
4.   **for each** player $p$ {
5.     InitializeToEmpty($h_p^{prev}$)
6.     InitializeToEmpty($h_p^{curr}$) }
7.   $APPE := true$
8.   $APS := true$
9.   $\delta := false$
10.   $t := 0$
11.   $\phi := \pi_{Me}^*$
12.   **while** $APS$ { // beginning of each epoch
13.     **repeat** $N^t$ **times** {
14.       Play($\phi$)
15.       **for each** player $p$
16.         Update($h_p^{curr}$) }
17.     **if** $APPE = false$ {
18.       **if** $\delta = false$
19.         **for each** player $p$
20.           **if** (Distance($h_p^{curr}, h_p^{prev}) > \epsilon_s^t$)
21.             $APS := false$
22.       $\delta := false$
23.       $a := \arg\max \text{V}(a, h_{-Me}^{curr})$
24.       **if** $\text{V}(a, h_{-Me}^{curr}) > \text{V}(\phi, h_{-Me}^{curr}) + n|A|\epsilon_s^{t+1}\mu$
25.         $\phi := a$ }
26.     **if** $APPE = true$
27.       **for each** player $p$
28.         **if** (Distance($h_p^{curr}, \pi_p^*) > \epsilon_e^t$) {
29.           $APPE := false$
30.           $\phi := \text{RandomAction}()$
31.           $\delta := true$ }
32.     **for each** player $p$ {
33.       $h_p^{prev} := h_p^{curr}$
34.       InitializeToEmpty($h_p^{curr}$) }
35.     $t := t + 1$ } }

We still need to discuss how to set the schedule for $(\epsilon_e^t, \epsilon_s^t, N^t)$. This is the topic of the next section.

### 3.5. Valid schedules

We now need to consider more precisely what good schedules are for changing the epochs' parameters. It turns out that the following conditions on the schedule for decreasing $\epsilon_e$ and $\epsilon_s$ while increasing $N$ are sufficient for the desirable properties to hold. The basic idea is to make $N$ go to infinity relatively fast compared to the $\epsilon_e$ and $\epsilon_s$. The reason for this exact definition will become clear from the proofs in the next section.

**Definition 4** *A schedule $\{(\epsilon_e^t, \epsilon_s^t, N^t)\}_{t \in \{0,1,2,...\}}$ is valid if*

- $\epsilon_s^t, \epsilon_e^t$ *decrease monotonically and converge to 0.*
- $N^t \to \infty$.
- $\prod_{t \in \{1,2,...\}}(1 - |A|_\Sigma \frac{1}{N^t(\epsilon_s^{t+1})^2}) > 0$ *(with all factors > 0), where $|A|_\Sigma$ is the total number of actions summed over all players.*
- $\prod_{t \in \{1,2,...\}}(1 - |A|_\Sigma \frac{1}{N^t(\epsilon_e^t)^2}) > 0$ *(with all factors > 0).*

The next theorem shows that a valid schedule always exists.

**Theorem 1** *A valid schedule always exists.*

**Proof**: Let $\{\epsilon_e^t = \epsilon_s^{t+1}\}_{t \in \{0,1,2,...\}}$ be any decreasing sequence going to 0. Then let $N^t = \left\lceil \frac{|A|_\Sigma}{(1 - \frac{1}{2^{(\frac{1}{t})^2}})(\epsilon_e^t)^2} \right\rceil$ (which indeed goes to infinity). Then, $\prod_{t \in \{1,2,...\}} 1 - |A|_\Sigma \frac{1}{N^t(\epsilon_s^{t+1})^2} = \prod_{t \in \{1,2,...\}} 1 - |A|_\Sigma \frac{1}{N^t(\epsilon_e^t)^2} \geq \prod_{t \in \{1,2,...\}} \frac{1}{2^{(\frac{1}{t})^2}}$ (we also observe that all factors are > 0). Also, $\prod_{t \in \{1,2,...\}} \frac{1}{2^{(\frac{1}{t})^2}} = 2^{\sum_{t \in \{1,2,...\}} \log \frac{1}{2^{(\frac{1}{t})^2}}} = 2^{\sum_{t \in \{1,2,...\}} -(\frac{1}{t})^2}$. Because the sum in the exponent converges, it follows that this is positive. ∎

## 4. AWESOME learns a best-response against eventually stationary opponents

In this section we show that if the other agents use fixed (potentially mixed) strategies, then AWESOME learns to play a best-response strategy against the opponents. This holds even if the opponents are nonstationary first (e.g., because they are learning themselves) as long as they become stationary at some time.

**Theorem 2** *With a valid schedule, if all the other players play fixed strategies forever after some round, AWESOME converges to a best response with probability 1.*

**Proof**: We prove this in two parts. First, we prove that after any given restart, with nonzero probability, the AWESOME player never restarts again. Second, we show that after any given restart, the probability of never restarting again without converging on the best response is 0. It follows that with probability 1, we will eventually converge.

To show that after any given restart, with nonzero probability, the AWESOME player never restarts again: consider the probability that for all $t$ ($t$ being set to 0 right after the restart), we have $\max_{p \neq 1}\{d(\phi_p^t, \phi_p)\} \leq \frac{\epsilon_s^{t+1}}{2}$ (where the AWESOME player is player 1, $\phi_p^t$ is the distribution of actions actually played by $p$ in epoch $t$, and $\phi_p$ is the (stationary) distribution that $p$ is actually playing from). This probability is given by $\prod_{t \in \{1,2,...\}}(1 - P(\max_{p \neq 1}\{d(\phi_p^t, \phi_p)\} > \frac{\epsilon_s^{t+1}}{2}))$, which is greater than $\prod_{t \in \{1,2,...\}}(1 - \sum_{p \neq 1} P(d(\phi_p^t, \phi_p) > \frac{\epsilon_s^{t+1}}{2}))$, which in turn is greater than $\prod_{t \in \{1,2,...\}}(1 - \sum_{p \neq 1} \sum_a P(|\phi_p^t(a) - \phi_p(a)| > \frac{\epsilon_s^{t+1}}{2}))$ (where $\phi_p(a)$ is the probability $\phi_p$ places on $a$). Because $E(\phi_p^t(a)) = \phi_p(a)$, and observing $Var(\phi_p^t(a)) \leq \frac{1}{4N^t}$, we can now apply Chebyshev's inequality and conclude that the whole product is greater than $\prod_{t \in \{1,2,...\}} 1 - |A|_\Sigma \frac{1}{N^t(\epsilon_s^{t+1})^2}$, where $|A|_\Sigma$ is the total number of actions summed over all players.[10] But for a valid schedule, this is greater than 0.

Now we show that if this event occurs, then $APS$ will not be set to *false* on account of the stationary players. This is because $d(\phi_p^t, \phi_p^{t-1}) > \epsilon_s^t \Rightarrow d(\phi_p^t, \phi_p) + d(\phi_p^{t-1}, \phi_p) > \epsilon_s^t \Rightarrow d(\phi_p^t, \phi_p) > \frac{\epsilon_s^t}{2} \lor d(\phi_p^{t-1}, \phi_p) > \frac{\epsilon_s^t}{2} \Rightarrow d(\phi_p^t, \phi_p) > \frac{\epsilon_s^{t+1}}{2} \lor d(\phi_p^{t-1}, \phi_p) > \frac{\epsilon_s^t}{2}$ (using the triangle inequality and the fact that the $\epsilon_s$ are stricly decreasing).

All that is left to show for this part is that, given that this happens, $APS$ will, with some nonzero probability, not be set to *false* on account of the AWESOME player. Certainly this will not be the case if $APPE$ remains *true* forever, so we can assume that this is set to *false* at some point. Then, with probability at least $\frac{1}{|A|}$, the first action $b$ that the AWESOME player will choose after $APPE$ is set to *false* is a best response to the stationary strategies. (We are making use of the fact that the stationary players' actions are independent of this choice.) We now claim that if this occurs, then $APS$ will not be set to *false* on account of the AWESOME player, because the AWESOME player will play $b$ forever. This is because the expected utility of playing any action $a$ against players who play from distributions $\phi_{>1}^t$ (call this $u_1(a, \phi_{>1}^t)$) can be shown to differ at most $n|A|\max_{p \neq 1} d(\phi_p, \phi_p^t)\mu$ from the expected utility of playing action $a$ against players who play from distributions $\phi_{>1}$ (call this $u_1(a, \phi_{>1})$). Thus, for any $t$ and any $a$, we have $u_1(a, \phi_{>1}^t) \leq$ $u_1(a, \phi_{>1}) + n|A|\epsilon_s^{t+1}\mu \leq u_1(b, \phi_{>1}) + n|A|\epsilon_s^{t+1}\mu$ (because $b$ is a best-response to $\phi_{>1}$), and it follows that the AWESOME player will never change its strategy.

---

[10]We observe that we used the fact that the schedule is valid to assume that the factors are greater than 0 in the manipulation.

Now, to show that after any given restart, the probability of never restarting again without converging on the best response is 0: there are two ways in which this could happen, namely with $APPE$ being set to *true* forever, or with it set to *false* at some point. In the first case, we can assume that the stationary players are not actually playing the precomputed equilibrium (because in this case, the AWESOME player would actually be best-responding forever). Let $p \neq 1$ and $a$ be such that $\phi_p(a) \neq \pi_p^*(a)$, where $\pi_p^*(a)$ is the equilibrium probability $p$ places on $a$. Let $d = |\phi_p(a) - \pi_p^*(a)|$. By Chebyshev's inequality, the probability that $\phi_p^t(a)$ is within $\frac{d}{2}$ of $\phi_p(a)$ is at least $1 - \frac{1}{N^t d^2}$, which goes to 1 as $t$ goes to infinity (because $N^t$ goes to infinity). Because $\epsilon_e^t$ goes to 0, at some point $\epsilon_e^t < \frac{d}{2}$, so $|\phi_p^t(a) - \phi_p(a)| < \frac{d}{2} \Rightarrow |\phi_p^t(a) - \pi_p^*(a)| > \epsilon_e^t$. With probability 1, this will be true for some $\phi_p^t(a)$, and at this point $APPE$ will be set to $false$. So the first case happens with probability 0. For the second case where $APPE$ is set to *false* at some point, we can assume that the AWESOME player is not playing any best-response $b$ forever from some point onwards, because in this case the AWESOME player would have converged on a best response. All we have to show is that from any epoch $t$ onwards, with probability 1, the AWESOME player will eventually switch actions (because starting at some epoch $t$, $\epsilon_s$ will be small enough that this will cause $APS$ to be set to *false*). If playing an action $a$ against the true profile $\phi_{>1}$ gives expected utility $k$ less than playing $b$, then by continuity, for some $\epsilon$, for any strategy profile $\phi'_{>1}$ within distance $\epsilon$ of the true profile $\phi_{>1}$, playing $a$ against $\phi'_{>1}$ gives expected utility at least $\frac{k}{2}$ less than playing $b$. By an argument similar to that made in the first case, the probability of $\phi_{>1}^t$ being within $\epsilon$ of the true profile $\phi_{>1}$ goes to 1 as $t$ goes to infinity; and because eventually, $n|A|\epsilon_s^{t+1}\mu$ will be smaller than $\frac{k}{2}$, this will cause the AWESOME player to change actions. ∎

## 5. AWESOME converges to a Nash equilibrium in self-play

In this section we show that AWESOME converges to a Nash equilibrium when all the other players are using AWESOME as well.

**Theorem 3** *With a valid schedule, AWESOME converges to a Nash equilibrium in self-play with probability 1.*

**Proof**: We first observe that the values of $APPE$ and $APS$ are always the same for all the (AWESOME) players, due to the synchronization efforts in the algorithm. It can be shown in a manner similar to the proof of Theorem 2 that after any restart, with nonzero probability, we have, for all $t$, $max_p\{d(\phi_p^t, \pi_p^*)\} \leq \epsilon_e^t$ (where $\phi_p^t$ is the distribution of actions actually played by $p$ in epoch $t$, and $\pi_p^*$ is the equilibrium strategy for $p$). In this case, $APPE$ is never set to *false* and the players play the equilibrium forever.

All that is left to show is that, after any restart, the probability of never restarting while not converging to an equilibrium is 0. This can only happen if $APPE$ is set to *false* at some point, and the players do not keep playing a pure-strategy equilibrium forever starting at some point after this. As in the proof of Theorem 2, all we have to show is that from any epoch $t$ onwards, with probability 1, some player will eventually switch actions (because starting at some epoch $t$, $\epsilon_s$ will be small enough that this will cause $APS$ to be set to false). Because we can assume that at least one player is not best-responding to the others' actions, the proof of this claim is exactly identical to that given in the proof of Theorem 2. ∎

It is interesting to observe that even in self-play, it is possible (with nonzero probability) that AWESOME players converge to an equilibrium other than the precomputed equilibrium. Consider a game with a pure-strategy equilibrium as well as a mixed-strategy equilibrium where every action is played with positive probability. If the mixed-strategy equilibrium is the one that is precomputed, it is possible that the equilibrium hypothesis (by chance) is rejected, and that each player (by chance) picks its pure-strategy action after this. Because from here on, the players will always be best-responding to what the others are doing, they will never change their strategies, the stationarity hypothesis will never be rejected, and we have converged on the pure-strategy equilibrium.

## 6. Conclusions

A satisfactory multiagent learning algorithm should, *at a minimum*, learn to play optimally against stationary opponents, and converge to a Nash equilibrium in self-play. Surprisingly, current algorithms, even those that specifically pursued this pair of properties as a goal, do not have these properties. The algorithm that has come closest is WoLF-IGA. It has been proven to have these two properties in simple 2-player 2-action games—further making the unrealistic assumptions that the opponent's strategy is known at each step, and using infinitesimally small gradient ascent steps.

In this paper we presented AWESOME, the first general algorithm that is guaranteed to learn to play optimally against opponents that are stationary (and against opponents that eventually become stationary), and to converge to a Nash equilibrium in self-play. It has these two desirable properties in all repeated games (with any finite number of agents, any finite number of actions, and unrestricted payoff matrices), requiring only that the other players' actual actions (not their strategies) can be observed at each step. AWESOME also does not use infinitesimal steps at any point of the algorithm.

The basic idea behind AWESOME (*Adapt When Everybody is Stationary, Otherwise Move to Equilibrium*) is to try to adapt to the other agents' strategies when they appear stationary, but otherwise to retreat to a precomputed equilibrium strategy.

At any point in time, AWESOME maintains either of two null hypotheses: that the others are playing the precomputed equilibrium, or that the others are stationary. Whenever both of these hypotheses are rejected, AWESOME restarts completely. AWESOME may reject either of these hypotheses based on actions played in an epoch. Over time, the epoch length is carefully increased and the criterion for hypothesis rejection tightened to obtain the convergence guarantee. The AWESOME algorithm is also self-aware: when it detects that its own actions signal nonstationarity to the others, it restarts itself for synchronization purposes.

## 7. Future research

The techniques used in proving the properties of AWESOME are fundamentally different from those used for previous algorithms, because the requirement that the opponents' whole strategies can be observed is dropped. These techniques may be valuable in the analysis of other learning algorithms in games.

The AWESOME algorithm itself can also serve as a stepping stone for future multiagent learning algorithm development. AWESOME can be viewed as a skeleton—that guarantees the satisfaction of the two minimal desirable properies—on top of which additional techniques may be used in order to guarantee further desirable properties.

There are several open research questions regarding AWESOME. First, it is important to determine which valid schedules give *fast* convergence. This can be studied from a theoretical angle, by deriving asymptotic running time bounds for families of schedules. It can also be studied experimentally for representative families of games. A related second question is whether there are any structural changes we can make to AWESOME to improve the convergence time while maintaining the properties derived in this paper. For instance, maybe AWESOME does not need to forget the entire history when it restarts. A third question is whether we can integrate learning the structure of the game more seamlessly into AWESOME (rather than first learning the structure of the game and then running AWESOME).

## References

Bowling, M., & Veloso, M. (2002). Multiagent learning using a variable learning rate. *Artificial Intelligence*, *136*, 215–250.

Cahn, A. (2000). *General procedures leading to correlated equilibria*. Discussion paper 216. Center for Rationality, The Hebrew University of Jerusalem, Israel.

Claus, C., & Boutilier, C. (1998). The dynamics of reinforcement learning in cooperative multiagent systems. *Proceedings of the National Conference on Artificial Intelligence (AAAI)* (pp. 746–752). Madison, WI.

Conitzer, V., & Sandholm, T. (2003). Complexity results about Nash equilibria. *Proceedings of the Eighteenth International Joint Conference on Artificial Intelligence (IJCAI)*. Acapulco, Mexico. Earlier version appeared as technical report CMU-CS-02-135.

Foster, D. P., & Vohra, R. V. (1997). Calibrated learning and correlated equilibrium. *Games and Economic Behavior*, *21*, 40–55.

Fudenberg, D., & Levine, D. (1998). *The theory of learning in games*. MIT Press.

Fudenberg, D., & Levine, D. (1999). Conditional universal consistency. *Games and Economic Behavior*, *29*, 104–130.

Hart, S., & Mas-Colell, A. (2000). A simple adaptive procedure leading to correlated equilibrium. *Econometrica*, *68*, 1127–1150.

Hu, J., & Wellman, M. P. (1998). Multiagent reinforcement learning: Theoretical framework and an algorithm. *International Conference on Machine Learning* (pp. 242–250).

Littman, M., & Stone, P. (2003). A polynomial-time Nash equilibrium algorithm for repeated games. *Proceedings of the ACM Conference on Electronic Commerce (ACM-EC)*. San Diego, CA.

Littman, M. L. (1994). Markov games as a framework for multi-agent reinforcement learning. *International Conference on Machine Learning* (pp. 157–163).

Papadimitriou, C. (2001). Algorithms, games and the Internet. *STOC* (pp. 749–753).

Sen, S., & Weiss, G. (1998). Learning in multiagent systems. In G. Weiss (Ed.), *Multiagent systems: A modern introduction to distributed artificial intelligence*, chapter 6, 259–298. MIT Press.

Singh, S., Kearns, M., & Mansour, Y. (2000). Nash convergence of gradient dynamics in general-sum games. *Proceedings of the Uncertainty in Artificial Intelligence Conference (UAI)* (pp. 541–548). Stanford, CA.

Tan, M. (1993). Multi-agent reinforcement learning: Independent vs. cooperative agents. *International Conference on Machine Learning* (pp. 330–337).

Vrieze, O. (1987). Stochastic games with finite state and action spaces. *CWI Tracts*.

Wang, X., & Sandholm, T. (2002). Reinforcement learning to play an optimal Nash equilibrium in team Markov games. *Proceedings of the Annual Conference on Neural Information Processing Systems (NIPS)*. Vancouver, Canada.